\documentclass[twocolumn,showpacs,preprintnumbers]{revtex4}
\usepackage{amsmath}
\usepackage{dcolumn}
\usepackage{bm}
\usepackage{epsfig}
\usepackage{epstopdf}
\usepackage{graphicx}
\usepackage{amsfonts}
\usepackage{amssymb}
\usepackage{mathrsfs}
\usepackage{color}
\usepackage{multirow}
\usepackage{appendix}

\begin{document}
\title{Single-atom demonstration of quantum Landauer principle}
\author{L. L. Yan$^{1}$}
\author{T. P. Xiong$^{1,2}$}
\author{K. Rehan$^{1,2}$}
\author{F. Zhou$^{1}$}
\email{zhoufei@wipm.ac.cn}
\author{D. F. Liang$^{1,3}$}
\author{L. Chen$^{1}$}
\author{J. Q. Zhang$^{1}$}
\author{W. L. Yang$^{1}$}
\email{ywl@wipm.ac.cn}
\author{Z. H. Ma$^{4}$}
\author{M. Feng$^{1,3,5,6}$}
\email{mangfeng@wipm.ac.cn}
\affiliation{$^{1}$ State Key Laboratory of Magnetic Resonance and Atomic and Molecular Physics,
Wuhan Institute of Physics and Mathematics, Chinese Academy of Sciences, Wuhan, 430071, China\\
$^{2}$ School of Physics, University of the Chinese Academy of Sciences, Beijing 100049, China \\
$^{3}$ Synergetic Innovation Center for Quantum Effects and Applications (SICQEA), Hunan Normal University,
Changsha 410081, China \\
$^{4}$ Department of Mathematics, Shanghai Jiaotong University, Shanghai, 200240, China \\
$^{5}$ Center for Cold Atom Physics, Chinese Academy of Sciences, Wuhan 430071, China \\
$^{6}$ Department of Physics, Zhejiang Normal University, Jinhua 321004, China }
\begin{abstract}
One of the outstanding challenges to information processing is the eloquent suppression of energy consumption in
execution of logic operations. Landauer principle sets an energy constraint in deletion of a classical bit of information.
Although some attempts have been paid to experimentally approach the
fundamental limit restricted by this principle, exploring Landauer principle in a purely quantum mechanical fashion is still
an open question. Employing a trapped ultracold ion, we experimentally demonstrate a quantum version of Landauer principle, i.e.,
an equality associated with energy cost of information erasure in conjunction with entropy change of the associated quantized environment.
Our experimental investigation substantiates an intimate
link between information thermodynamics and quantum candidate systems for information processing.
\end{abstract}
\maketitle

It was Rolf Landauer who expostulated, for the first time, a minimum amount of energy required to
be consumed in deletion of a classical bit of information, known as Landauer principle (LP) \cite{landauer}, implying an
irreversibility of logical operations \cite {landauer1,lloyd,meindl}.
In terms of the LP, the erasure of information is fundamentally a dissipative process,
which dissipates at least $k_{B}T\ln 2$ amount of heat, called Landauer
bound, from the system into the attached reservoir, where $k_{B}$
and $T$ represent Boltzmann constant and  reservoir temperature,
respectively. On the other hand, if we try to understand
violation of the second law of thermodynamics by the Maxwell's
demon (an intellectual creature envisioned by Maxwell), who converts thermal energy of the
reservoir into useful work \cite{penrose,bennett,demon},
entropy cost should be considered regarding the demon's
memory, which is also subject to the LP. So far, much of discussion has
been devoted to the validity and usefulness of the LP, including some
skepticism and misunderstanding \cite{mar,argue0,argue1,argue2,argue3}. Since
suppression of energy dissipation to a possible minimum is indispensable towards
the continued development of digital computers \cite{frank,pop}, even for
quantum information processing \cite{plenio}, further inquisition of the LP is certainly of
peculiar importance.

\begin{figure*}[htbp]
\centering
\includegraphics[width=16 cm,height=7.3 cm]{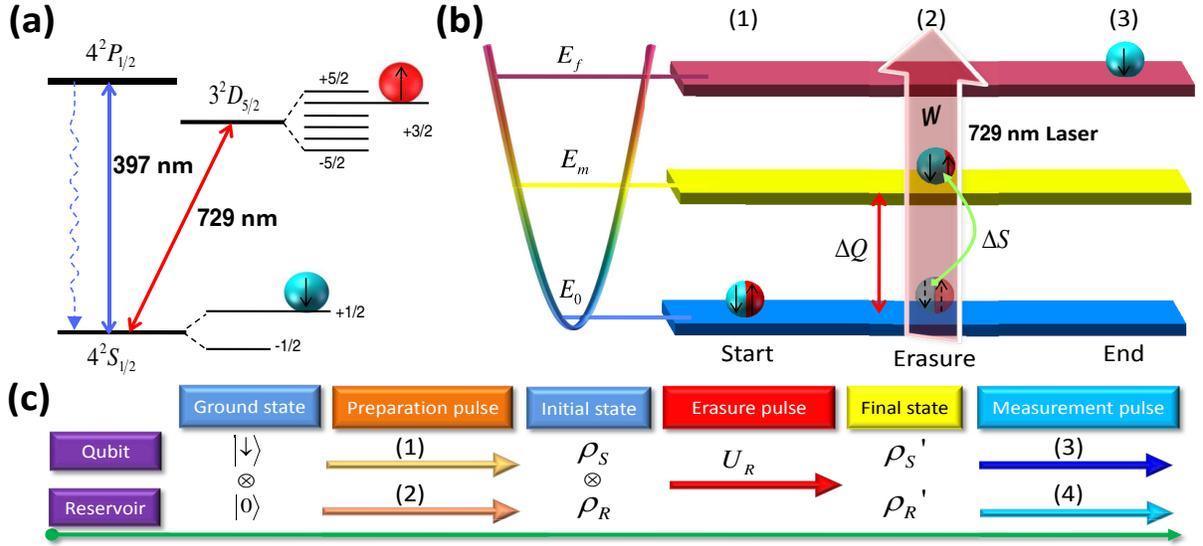}
\caption{A single trapped ion manipulated for exploring an improved LP. (a) Level scheme of the $^{40}Ca^+$ ion confined in a linear Paul trap under an external magnetic field. The qubit encoded in $\mid\uparrow\rangle$ and $\mid\downarrow\rangle$ represents the system and the reservoir is denoted by the quantized $z$-axis vibration of the ion. The system-reservoir coupling is well controlled by red- or blue-sideband 729-nm laser radiation, and readout of the qubit is carried out by 397-nm spontaneous emission. (b) Illustrative schematic. (1) Starting point, where the qubit (i.e., the system) is in a maximally mixed state $\rho_S=(\mid\uparrow\rangle\langle\uparrow\mid + \mid\downarrow\rangle\langle\downarrow\mid)/2$ and the vibrational degree of freedom (i.e., the reservoir) is thermalized to energy $E_{0}$. (2) Mediate step, where the 729-nm laser drives a unitary evolution of the system and reservoir. The entropy of the system decreases, while the energy of the reservoir increases to $E_m$. (3) End of the erasure, where the system is populated in $\mid\downarrow\rangle$ and the reservoir ends at a higher level with energy $E_f$. (c) Flow-diagram of the experimental operations, where (1) is to reach a mixed state of the system; (2) is related to a thermal state of the reservoir; (3) is to monitor the population of the system state and (4) refers to measuring the average phonon number in the reservoir by observing the blue-sideband Rabi oscillation.}
\label{Fig1}
\end{figure*}

Some experimental attempts, subject to the conventional inequality of the LP, have
been carried out to approach the Landauer bound using single-bit
operations \cite{exp0,exp00,exp1,exp2,exp3}.
The latest experiment \cite{exp3}, for example, was implemented on a nanoscale digital magnetic memory
bit, by which the intrinsic energy dissipation per single-bit operation was detected.
In contrast, a very relevant investigation
for extracting $k_{B}T\ln 2$ of heat in creation of one-bit of
information \cite{koski}, which is the complementary study of the LP in
the context of a Szil{\'a}rd engine \cite{engine}-a quantitative Maxwell's demon, was accomplished
recently on a single-electron transistor. However, all these
aforementioned considerations surmise that the heat related to removal or
generation of a bit of information is regarding an open environment,
which is a reservoir with much bigger size than the system. If we
extend the treatments to a quantum domain, e.g., the
information to be erased being encoded in a qubit and the environment being quantized, the model
needs to be substantially reconsidered. The entropy of a quantum system is associated with the
characteristic of the state, rather than the thermodynamic arrow of time.
Besides, in contrary to a large reservoir considered in the original version of the LP, the quantized reservoir is of a finite size
and hence vacillating in interaction with the qubit system under quantum operations for information erasure. So, different from
the previous assumption of final product states \cite{pie00,lr03,qse}, quantum
correlation appears between the system and the quantized reservoir during a quantum
erasure process, which requires evaluation via mutual information
and relative entropy. However, even with elaborately designed systems under the consideration of a quantized reservoir, experimentally
measuring mutual information and relative entropy is not easy
to accomplish \cite{q-exp-1,q-exp-2}, which hampers the efforts to exactly approach the Landauer bound.

Here we report an experimental investigation of a quantum mechanical LP by an experimental evaluation of
system-reservoir correlation and entropy change during the erasure process. Our operations are based on a single ultracold
$^{40}Ca^{+}$ ion confined in a linear Paul trap. Trapped ultracold ions, with the possibility of precise
manipulation, have been considered as an ideal platform to
explore the thermodynamics in quantum domain with ultimate accuracy
\cite{quant-exp,ion-thermo-1,ion-thermo-2,ion-thermo-3}. For our purpose, we consider the two
internal levels of the ion as the qubit system and the vibrational
degree of freedom of the ion as a finite-temperature reservoir.
Removing information encoded initially in the qubit, we
perceive the quantized LP by observing the phonon number variation
in the quantized vibration of the ion. Owing to precise control of
the two degrees of freedom in a coherent way, we demonstrate for the first time
the possibility to approach the Landauer bound solely in a quantum mechanical
fashion.

Our exploration mainly follows the idea termed as an improved LP \cite{newprinciple}, which describes the energy cost regarding information erasure employing an
equality, rather than the conventional inequality. Assuming initial states of the system and reservoir
to be a mixed state $\rho_{S}$ and a thermal state $\rho_{R}$ correspondingly, our model starts from an uncorrelated state
$\rho_{SR}=\rho_{S}\otimes\rho_{R}$ and evolves to a system-reservoir
correlated state $\rho'_{SR}$ resulted from an erasure process, where
the system and reservoir states turn to $\rho'_{S}=\text{Tr}_{R}[\rho'_{SR}]$
and $\rho'_{R}=\text{Tr}_{S}[\rho'_{SR}]$, respectively. The improved LP gives
a tight bound of the heat cost for an information erasure by an equality as below,
\begin{equation}
\Delta Q/k_{B}T = \Delta S+I(S^{\prime}\colon R^{\prime})+D(\rho^{\prime}_R\Vert\rho_R),
\label{eq1}
\end{equation}
where $\Delta Q=\text{Tr}[H_R\rho'_{R}]-\text{Tr}[H_R\rho_{R}]$ represents an average energy increase of the thermal reservoir with the Hamiltonian $H_R$. $\Delta
S=S(\rho_{S})-S(\rho'_{S})$ denotes the entropy decrease of the system with the Von Neumann entropy defined as $S(\hat{o})=-\text{Tr}[\hat{o}\log\hat{o}]$. $I(S^{\prime}\colon R^{\prime})=S(\rho^{\prime}_{S})+S(\rho^{\prime}_{R})-S(\rho^{\prime}_{SR})$ indicates mutual information related to the system-reservoir
correlations following the erasure process, and the relative entropy
$D(\rho^{\prime}_R\Vert\rho_R)=\text{Tr}[\rho^{\prime}_R\log\rho^{\prime}_R]-\text{Tr}[\rho^{\prime}_R\log\rho_R]$
corresponds to the free energy increase of the reservoir during the
erasure process. In comparison with the original form of the LP, i.e., $\Delta Q/k_{B}T \ge\Delta S$, Eq. (\ref{eq1}) is obviously a quantum
version of the LP linking quantum information with thermodynamics.

\begin{figure*}[htbp]
\centering
\includegraphics[width=14.0cm,height=8.8 cm]{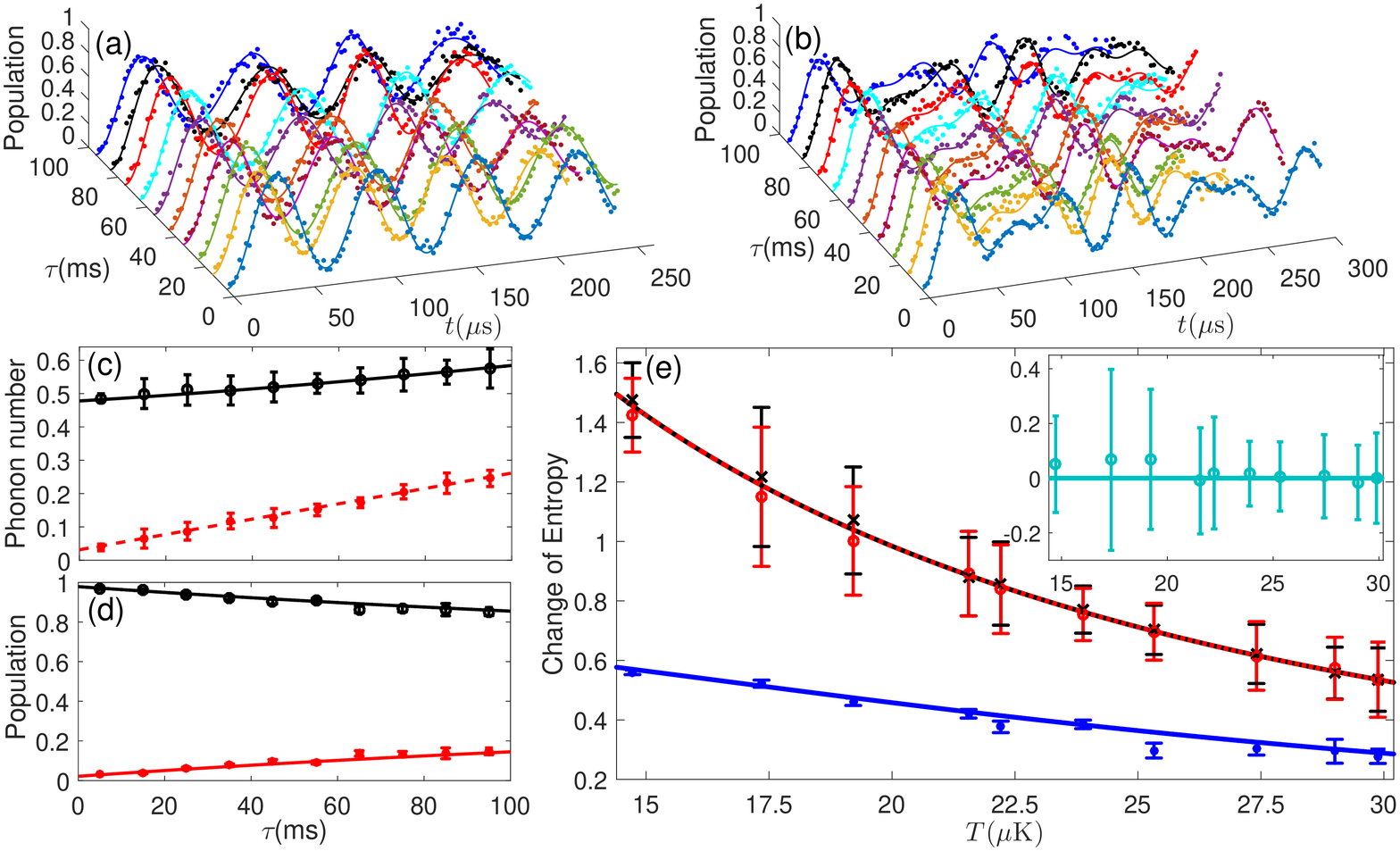}
\caption{Experimental implementation of information erasure in a qubit system. (a) Blue-sideband Rabi oscillations before the erasure for different waiting time $\tau$. (b) Blue-sideband Rabi oscillation at the end of erasure for different waiting time $\tau$. (c) Average phonon number in the reservoir before the erasure (red dots) and after the erasure (black circlets), obtained by calculating the data in (a) and (b). (d) Population in $\mid\downarrow\rangle$ (black circlets) and $\mid\uparrow\rangle$ (red dots) at the end of erasure. The system is initialized with populations of $0.531(15)$ and $0.467(16)$ in $\mid\downarrow\rangle$ and $\mid\uparrow\rangle$, respectively. Dots or circlets are experimental data and curves are analytical results. Each data in (a) and (b) is measured by $10^2$ repetition and in (d) by $10^4$ repetition. (e) Test of the improved LP at different reservoir temperatures. The curves from top to bottom correspond to $\Delta Q/k_{B}T$ with $\Delta Q=Q_0(\langle n\rangle-\langle n\rangle_0)$ (black curve and crosses), $\Delta S + I(S^{\prime}\colon R^{\prime})+D(\rho^{\prime}_R\Vert\rho_R)$ (red curve and circlets) and $\Delta S$ (blue curve and dots), where the reservoir temperature is defined as $ T=T_0/\ln(1+1/\langle n\rangle_0)$ with $\langle n\rangle_0$ the initial average phonon number. Inset of (e) demonstrates the result of $\Delta Q/k_{B}T -[ \Delta S+I(S^{\prime}\colon R^{\prime})+D(\rho^{\prime}_R\Vert\rho_R)]$. Experimental data in (e) are obtained from panels (a-d). The Energy and temperature units are $Q_0:=\hbar\omega_z=1.7\times 10^{-28}$ J and $T_0:=Q_0/k_B=48.5 $ $\mu$K, respectively. The curves in each panel are analytical results in our optimal erasure scheme using a $\pi$ pulse of the red-sideband radiation \cite{sm}. The associated error bars reveal standard deviation. } \label{Fig2}
\end{figure*}

Our experiment is performed in a typical linear Paul trap with four parallel blade-like electrodes, and two end-cap electrodes
aligned along the axial direction \cite{sa,sm}. The trap is driven by the rf-frequency $\Omega_{rf}/2\pi = $17.6 MHz with the power of 5 W, giving the radial frequency
$\omega_r/2\pi=1.2$ MHz, and the trap axial frequency under the pseudopotential approximation is
$\omega_z/2\pi=1.01$ MHz with a voltage of 720 V applied to the end-cap electrodes.
Under an external magnetic field of 6 Gauss, the
ground state $4^2S_{1/2}$ and the metastable state $3^2D_{5/2}$
split into two and six hyperfine energy levels, respectively, as
shown in Fig. \ref{Fig1}(a). We encode
$\mid\downarrow\rangle$ in $|4^{2}S_{1/2}, m_{J}=+1/2\rangle$ and
$\mid\uparrow\rangle$ in $|3^{2}D_{5/2}, m_{J}=+3/2\rangle$, where
$m_{J}$ is the magnetic quantum number. After Doppler cooling and
optical pumping, a resolved sideband cooling is applied to
cool the $z$-axis motional mode down to its vibrational ground state
with the final average phonon number of $0.030(7)$ and the
internal state of the ion is initialized to $\mid\downarrow\rangle$ with
a probability of 98.9(2.3)\%. The population of $\mid\downarrow\rangle$ could be detected by a 397-nm laser with detecting error $\le 0.22(8)\%$ \cite{xiong}.
A narrow-linewidth 729-nm laser with wave
vector $\bm{k}$ radiates the ultracold ion with an incident angle of
$22.5^{\circ} $ with respect to the $z$-axis, yielding the Lamb-Dicke
parameter $\eta$ of 0.09.

Figure \ref{Fig1}(b) presents a schematic of our work. We start from preparing the qubit in an
equal probability admixture of the states $\mid\downarrow\rangle$ and
$\mid\uparrow\rangle$, indicating that the initial entropy of the system
is superlative but in contrast our available information about the
system is least, that is, we have no explicit information about the existing state of
the ion. For a general case, we consider the reservoir to be
initially in a thermal state with an associated energy
defined as $E_0=\text{Tr}[H_{R}\rho_R]$ for a given state $\rho_R$ (see Fig. \ref{Fig1}(b1)).
The erasure is performed by employing a red-sideband transition, governed by
$H_{R}=\eta\hbar\Omega(a\sigma_+e^{i\phi}+a^{\dagger}\sigma_-e^{-i\phi})/2$,
where $\Omega$ and $\phi$ represent the Rabi frequency and laser
phase, respectively, $\sigma_{+}$ ($\sigma_{-}$) indicates the
raising (lowering) operator of the spin, and $a$($a^{\dagger}$)
denotes the annihilation (creation) operator of the quantized
vibration of the ion in the $z$-direction. The erasure process
produced by $U_{R}=\exp(-iH_{R}t)$, with $t$ as the pulse length,
causes the population transfer from $\mid\uparrow\rangle$
to $\mid\downarrow\rangle$, corresponding to an entropy
decrease of the system which is accompanied by the
reservoir energy increase up to $E_m$, as depicted in Fig.
\ref{Fig1}(b2). The erasure process ends with the system in
the polarized state $\mid\downarrow\rangle$ and the reservoir changes
to the final state $\rho_R^{\prime}$, energy of which is given as
$E_f=\text{Tr}[H_R\rho^{\prime}_R]$, see Fig. \ref{Fig1}(b3).
So we finally have a complete enlightenment about the state of the system
accomplished by virtue of the heat increment $\Delta
Q=E_f-E_0$. In this way, we may verify Eq. (\ref{eq1}) by observing
the variation of population and phonon number in the qubit and the reservoir,
respectively, where the entropy decrease of a system is $\Delta S=\ln$2 in an ideal
erasure for the system attached to a zero-temperature reservoir. The corresponding experimental procedure is drawn in Fig. \ref{Fig1}(c).

\begin{figure}
\centering
\includegraphics[width=9.5cm,height=7.0 cm]{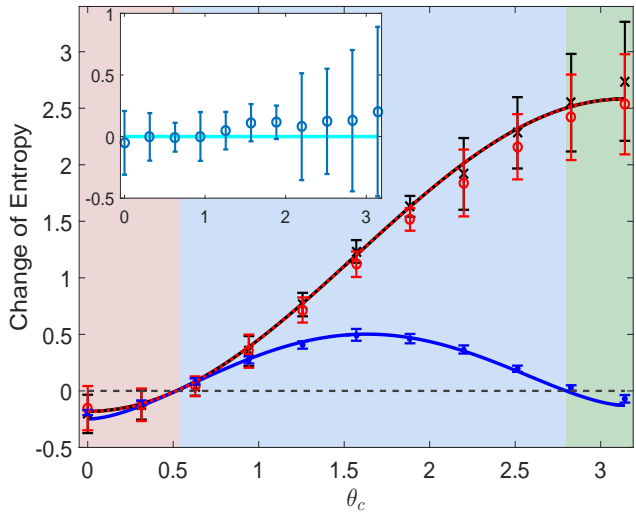}
\caption{Experimental test of the improved LP under different initial states of the system. Curves from top to bottom correspond to $\Delta Q/k_{B}T$ (black curve and crosses), $\Delta S+ I(S^{\prime}\colon R^{\prime})+D(\rho^{\prime}_R\Vert\rho_R)$ (red curve and circlets) and $\Delta S$ (blue curve and dots), where the reservoir temperature is initially $18.2(8)$ $\mu$K corresponding to the average phonon number $\langle n\rangle_0=$0.074(9). The horizontal dashed line indicates the positive-negative switch of the entropy decrease with boundary points at $\theta_c=0.54(3)$ and $2.80(2)$. Inset demonstrates the result of $\Delta Q/k_{B}T -[ \Delta S+I(S^{\prime}\colon R^{\prime})+D(\rho^{\prime}_R\Vert\rho_R)]$. Dots, circlets and crosses represent  experimental data; Curves are from analytical results. The system entropy is initially $S_1=S(\text{diag}(\cos^2(\theta_c/2),\sin^2(\theta_c/2)))$ with diag indicating a diagonal matrix and $\theta_c$ associated with the initial state of the system. The error bars indicate standard deviation.}
\label{Fig3}
\end{figure}

In our actual implementation, superposition of the qubit states was achieved by a carrier-transition operator $U_{C}(\theta_c,0) = \cos(\theta_c/2)I -i\sin(\theta_c/2)\sigma_{x}$, where $\sigma_x$ is the Pauli operator. Since the dephasing rate ($\sim$0.5 kHz) of the qubit is much larger than its decay rate ($\sim$1 Hz), waiting for a time longer than 2 ms would definitely lead to qubit dephasing but without affecting the population, which guarantees the removal of the off-diagonal terms of density matrix of the aforementioned model while retaining the diagonal terms unaffected, yielding the initial state $\rho_S=\alpha\mid\downarrow\rangle\langle\downarrow\mid+\beta\mid\uparrow\rangle\langle\uparrow\mid$ with $\alpha=\cos^2(\theta_{c}/2)$ and $\beta=\sin^2(\theta_{c}/2)$. Meanwhile, the reservoir was thermalized due to switch-off of the cooling lasers during this waiting time. Consequently, our implementation commences with the reservoir at a finite temperature, where the average phonon number $\langle n\rangle_{0}$ is determined by waiting for a desired time duration. To remove the information from the system as
much as we can, we have considered an optimal erasure under the first-order approximation where the largest contribution term affecting the erasure is eliminated during the red-detuned laser radiation for $t_{\text{op}}=\pi/\eta\Omega=$ 33(2) $\mu$s \cite{sm}. With this short time implementation, decoherence effects, during the information removal, are negligible.

In our current study, all the four parts of Eq. (1) are explored experimentally as below. In our setup, the knowledge about the phonon number is obtained from qubit state measurement under the government of the blue-detuning hamiltonian $H_B=\eta\hbar\Omega(a\sigma_- e^{i\phi}+a^{\dagger}\sigma_+ e^{-i\phi})/2$. Shown in Fig. \ref{Fig2}(a,b) are, respectively, the experimental observations of the blue-sideband Rabi oscillations versus the waiting time before and after the erasure, which is accomplished via curve
 fitting approach \cite{ra}. Consequently, based on the experimental data, the phonon number due to heat variation, as plotted in Fig. \ref{Fig2}(c), could be calculated by Eq. (S14) in \cite{sm}, and meanwhile the relative entropy of the reservoir is acquired.
However, the populations in $\mid\downarrow\rangle$ and $\mid\uparrow\rangle$ at the end of the erasure are monitored from the 397-nm spontaneous-emission fluorescence spectrum (see Fig. \ref{Fig2}(d)), which yields variation of the system entropy. Moreover, the mutual information term is smaller with respect to other terms in Eq. (\ref{eq1}), and a direct measurement of the mutual information is complicated due to its relevance to the system-reservoir entanglement. Therefore, we estimated the mutual information by an alternative way, as explained in \cite{sm}, instead of direct experimental measurements.

With the data measured above, the Landauer bound and the improved LP are witnessed under the different reservoir temperatures in Fig. \ref{Fig2}(e).
Our observations clearly exhibit that the equality in the original form of the LP cannot be accessed quantum mechanically at the finite temperature. In contrary to the classical version of the LP in which the huge reservoir changes very little during the erasure process such that the mutual information and the relative entropy of the reservoir can be ignored, both states of the system and reservoir in quantum regime vary during the erasure process, which yields a quantum correlation between the system and the reservoir. Intrinsically, we can observe a large disparity between the heat consumption of the reservoir (black curves) and the entropy decrease of the system (blue curves), due to the difference from quantum correlation and the relative entropy of reservoir.
This difference turns out to be larger at lower temperature because in this case the relative entropy becomes more dominant in Eq. (\ref{eq1}). Especially, in the limit of zero temperature, one may find the system's entropy decrease $\Delta S=\ln2$, the mutual information $I(S':R')=0$ and the relative entropy $D(\rho^{\prime}_R\Vert\rho_R)\rightarrow\infty$, which agrees with the result of $1/T\rightarrow\infty$ in the left-hand side of Eq. (\ref{eq1}). In this scenario, the nearly perfect overlap between the red dashed and black solid curves implies that Eq. (\ref{eq1}) provides a better way to understand the LP and the associated Landauer bound in quantum regime.

It is interesting to note that the quantum LP is very sensitive to initial condition of the model. Figure \ref{Fig3} presents the variation of the system entropy from the negative to positive and then to negative again, arguing that entropy decrease, i.e., removal of information, occurs in the system for $\theta_{c}\in$ [0.54(3), 2.80(2)], and for else, the system entropy increases during the erasure indicating creation of information. This indicates that our implementation could demonstrate both the generation and deletion of the system's information by simply tuning $\theta_c$ (i.e., varying the initial state of the system). Particularly, the generation of information in the region of $\theta_{c}<$0.54(3) corresponds to a fully quantized single-qubit Szil{\'a}rd engine, the opposite process of the quantum LP. However, the observation in the region of $\theta_{c}>$2.80(2) is counter-intuitive, where the generation of information is accompanied by the energy increase of the reservoir. This phenomenon is completely different from the aforementioned quantum LP and Szil{\'a}rd engine, but reflects the significant role of the relative entropy of the reservoir played in the fully quantized system-reservoir model \cite{vedral}.


In summary, we have demonstrated in a more pertinent way, the first experimental investigation of the quantum mechanical LP using a single ultracold trapped ion. This is an imperative step towards  better understanding of the fundamental physical limitations of irreversible logic operations at quantum level. Our observation confirms that the LP still holds even at quantum level after some modification by introducing quantum information quantities. Our experimental evidence might be helpful for efficient initialization of a future quantum computer involving an artificial quantum reservoir
for fast eliminating the encoded information from large numbers of qubits, in which the information erasure yields more heat consumption than the classical counterpart due to system-reservoir correlations. For the same reason, quantum error-correction in such a quantum computer would also cost higher heat (or energy).
We believe that the experiment reported here will open an avenue towards further exploration of the Landauer bound in quantum regime as well as the possible applications.

This work was supported by National Key Research $\&$ Development Program of China under grant No. 2017YFA0304503, by National Natural Science Foundation of
China under Grant Nos. 11734018, 11674360, 11404377, 91421111 and 11371247, and by the Strategic Priority Research Program of the Chinese
Academy of Sciences under Grant No. XDB21010100. K.R. acknowledges thankfully support from CAS-TWAS president's fellowship. LLY, TPX and KR contributed equally to this work.

\end{document}